\documentclass[aps,pra,twocolumn,showpacs]{revtex4}
\usepackage{amsmath, amssymb}

\newtheorem{observation}{Observation}
\newtheorem{proposition}{Proposition}

\newtheorem{definition}{Definition}
\newtheorem{theorem}{Theorem}

\begin{document}
\title[The Schmidt number as a universal entanglement measure]{The Schmidt number as a universal entanglement measure}

\author{J. Sperling}\affiliation{Arbeitsgruppe Quantenoptik, Institut f\"ur Physik, Universit\"at Rostock, D-18051 Rostock, Germany}\email{jan.sperling2@uni-rostock.de}
\author{W. Vogel}\affiliation{Arbeitsgruppe Quantenoptik, Institut f\"ur Physik, Universit\"at Rostock, D-18051 Rostock, Germany}\email{werner.vogel@uni-rostock.de}

\date{August 27, 2009}

\begin{abstract}
The class of local invertible operations is defined and the invariance of entanglement under such operations is established.
For the quantification of entanglement, universal entanglement measures are defined, which are invariant under local invertible transformations.
They quantify entanglement in a very general sense.
It is shown that the Schmidt number is a universal entanglement measure, which is most important for the general amount of entanglement.
For special applications, pseudo-measures are defined to quantify the useful entanglement for a certain quantum task.
The entanglement quantification is further specified by operational measures, which include the accessible observables by a given experimental setup.
\end{abstract}
\maketitle

\section{Introduction}

Entanglement is the key resource of the vast fields of Quantum Information Processing, Quantum Computation, and Quantum Technology, for an introduction see e.g.~\cite{book1,book2}.
For example, applications of entangled states are those for quantum key distribution~\cite{ekert91}, quantum dense coding~\cite{bennett-wiesner92}, and quantum teleportation~\cite{bennett93}.
Thus both the identification and the quantification of entanglement play a mayor role for future applications~\cite{book3}.

The phenomenon entanglement is closely related to the superposition principle of quantum mechanics.
A pure separable state is represented by a product of states for both systems.
A general pure state is a superposition of factorizable states.
The number of superpositions of factorizable states is given by the Schmidt rank~\cite{book1}.
A separable mixed quantum state is a convex combination of pure factorizable quantum states~\cite{Werner}.
The generalization of the Schmidt rank to mixed quantum states delivers the Schmidt number.
This generalization and the introduction of Schmidt number witnesses is given in~\cite{Sanpera2,Terhal,Bruss}.
The Schmidt number of a mixed quantum state fulfills the axioms of an entanglement measure, cf.~\cite{Vedral,Vidal2,Vedral2}.
More precisely, it is a convex roof measure as defined in~\cite{Uhlmann,Bennett}.

Since the amount of entanglement cannot increase under local operations and classical communication, all entanglement measures must satisfies the local operations and classical communication (LOCC) paradigm.
But in general, different entanglement measures do not deliver the same ordering of entangled quantum states~\cite{EisertPlenio,Miranowicz,Measures}.
It is important to note that for a given quantum task an adequate definition of the corresponding LOCC plays a crucial rule for the entanglement quantification.

Maximally entangled states are usually considered to have the highest amount of entanglement.
It turns out that for different quantum tasks different entangled quantum states are beneficial~\cite{GrossEisert}.
The experimentally demonstrated possibility of noise-free linear amplification makes a larger class of separable operations accessible~\cite{Ralph}.
Recently, it has been shown that these separable operations can increase entanglement with respect to certain measures~\cite{Duan}.
A subclass of separable operations -- so-called local filter operations -- play a crucial role for universal entanglement distillation protocol~\cite{HorodeckiEtAl}.
This necessitates a careful consideration of local invertible transformations in the context of entanglement measures.
It is also important that all entangled quantum states have some usable amount of entanglement, non-locality, and potential applications for quantum processing~\cite{Masanes1,Masanes2,Masanes3}.
The inclusion of all these aspects requires a critical study of the entanglement quantification.

In the present contribution we study entanglement measures and especially the Schmidt number.
We discuss the notion of maximally entangled states in connection with an arbitrary entanglement measure.
We conclude that in the most general sense the available amount of entanglement depends on the Schmidt number.
We also define pseudo-measures and operational measures, which quantify the entanglement for a specific quantum task and for a special experimental setup, respectively.

The paper is structured as follows.
In Sec.~\ref{Sec:EM} we discuss general entanglement measures and their properties in relation to one special entanglement measure -- the Schmidt number.
The definition of entanglement pseudo-measures and their application to an arbitrary experimental situation is given in Sec.~\ref{Sec:PM}.
A summary and some conclusions are given in Sec.~\ref{Sec:SC}.

\section{Entanglement Measures}\label{Sec:EM}
In this section we consider special properties of entanglement measures in connection with LOCC.
Starting with the given definitions of entanglement measures we obtain some properties which indicate a fundamental role of the Schmidt number.
Here and in the following we assume finite, but arbitrary dimensional Hilbert spaces $\mathcal{H}=\mathcal{H}_1\otimes\mathcal{H}_2$.
The generalization for continuous variable entanglement directly follows from the method of finite spaces as presented in Ref.~\cite{SpeVo2}.

\subsection{The LOCC Paradigm and Entanglement Measures}
Entanglement measures are usually defined by using LOCC.
Therefore an accurate definition of these operations is essential for understanding entanglement measures.
The general idea of a LOCC is a map that cannot create entanglement.
Therefore let us write the most general form of such a map $\Lambda$, cf.~\cite{book2,Vedral2,Rains}, the separable operations
$\Lambda\in\mathcal C_{\rm sep}$:
\begin{align}
	\Lambda(\rho)&=\sum_i[A_i\otimes B_i]\rho[A_i\otimes B_i]^\dagger.
\end{align}
The quantum state will be normalized by
\begin{align}
	\rho&\mapsto\rho'=\frac{\Lambda(\rho)}{{\rm tr}\,\Lambda(\rho)}.
\end{align}
The operations in the set $\mathcal C_{\rm sep}$ are also called stochastic separable operations.
An important subset is $\mathcal C_{\rm LU}$, which denotes all local unitaries $[U_1\otimes U_2]\rho[U_1\otimes U_2]^\dagger$.

A more general substructure $\mathcal C_{\mathcal X}$ of $\mathcal C_{\rm sep}$ is defined in the following way.
We have a set of operations $\mathcal X\subset\mathcal C_{\rm sep}$.
The substructure is defined by all $\Lambda\in\mathcal X$ and all compositions of elements of $\mathcal X$,
\begin{align}\label{Eq:SubSemiGroup}
	\Lambda_1,\Lambda_2\in\mathcal C_{\mathcal X} \quad\Rightarrow\quad \Lambda_1\circ\Lambda_2\in\mathcal C_{\mathcal X}.
\end{align}
This algebraic structure of a semi-group.
It is usually considered that this substructure must at least include local changes of the basis by local unitaries $\mathcal C_{\rm LU}$.
Here and in the following, we will call such a substructure, $\mathcal C_{\mathcal X}$ LOCC.
These particular LOCC are the applied operations for a special quantum task, for example quantum key distribution~\cite{ekert91}, or quantum dense coding~\cite{bennett-wiesner92}.

The usually considered class of LOCC operations is given by two-way classical communications, cf. \cite{book2}, for quantum teleportation~\cite{bennett93}.
These operations define a substructure of stochastic separable operations as given in Eq.~(\ref{Eq:SubSemiGroup}).
However, for other processes like distillation protocols the whole set of stochastic separable operations are applied~\cite{HorodeckiEtAl,Rains}.

For further studies, we may consider operations of the Form $A\otimes B$, so-called local filtering operations~\cite{book2,HorodeckiEtAl}, with
\begin{align}
	\nonumber \Lambda(|a\rangle\langle a|\otimes|b\rangle\langle b|)=&\Lambda_A(|a\rangle\langle a|)\otimes\Lambda_B(|b\rangle\langle b|)\\
	=&A|a\rangle\langle a|A^\dagger \otimes B|b\rangle\langle b|B^\dagger.
\end{align}
Note that, the problem of normalization, $A^\dagger A\otimes B^\dagger B\leq \mathbb I\otimes \mathbb I$, is discussed in~\cite{Measures}.
Beside such substructures we define the following LOCC.

\paragraph{$\mathcal C_{\rm LI}$, local invariables:}
Operations of the form
\begin{align}
	\rho \mapsto [T_1\otimes T_2]\rho[T_1\otimes T_2]^\dagger,
\end{align}
with $T_1$ and $T_2$ invertible, are called local invertibles.
Note that for any operation exists an inverse operation given by $T_1^{-1}$ and $T_2^{-1}$.
Further on, the LOCC $C_{\rm LI}$ include the identity, $\mathbb I_1\otimes\mathbb I_2$.
Thus, $C_{\rm LI}$ is a group.
A subgroup of these operations is $\mathcal C_{\rm LU}$.

\paragraph{$\mathcal C_{\rm LP}$, local Projections:}
In addition to all local unitaries this set contains local projections given by
\begin{align}
	\rho \mapsto [P_1\otimes P_2]\rho[P_1\otimes P_2],
\end{align}
with $P_1$ and $P_2$ projection operators.
Recently, we used these operators to show that continuous variable entanglement can always be identified in finite dimensions~\cite{SpeVo2}.

Now we can define a general entanglement measure $E$, which must fulfill the following definition, see~\cite{book3}.
\begin{definition}\label{Def:Measure}
	$E$ is an entanglement measure, if:
	\begin{align*}
		&{\rm (i)}\quad \sigma \mbox{ \rm separable} \quad \Leftrightarrow \quad E(\sigma)=0,\\
		&{\rm (ii)}\quad \forall \Lambda\in\mathcal C_{\mathcal X}: E(\rho)\geq E\left(\frac{\Lambda(\rho)}{{\rm tr}\,\Lambda(\rho)}\right).
	\end{align*}
\end{definition}
Usually, a third condition is that an entanglement measure must be invariant under local unitaries.
However, Condition~(ii) implies that the additional invariance under local unitaries is superfluous, see~Appendix~\ref{App:LU}.
Note that, instead of Condition~(ii) often a non-increasing behavior on average of the entanglement measure is considered, cf. our related comments in Sec.~\ref{SSec:mixed}.

Last but not least, Definition~\ref{Def:Measure} depends on the chosen LOCC.
Thus the precise mathematical definition of the LOCC $\mathcal C_{\mathcal X}$ -- as given above -- is crucial for the measure itself.
The physical interpretation of LOCC is given in Sec.~\ref{Sec:PM}.
We discussed above that different quantum tasks use different sets of LOCC.
For example, quantum teleportation uses deterministic two-way classical communication, and distillation uses all stochastic separable operations.

\subsection{Example: Schmidt Number}
A well-known example for an entanglement measure is the Schmidt number $r_{\rm S}$.
Let us consider the pure state $|\Psi\rangle$ with a Schmidt decomposition~\cite{book1},
\begin{align}\label{SchmDec}
	|\Psi\rangle=\sum_{n=1}^{r(\Psi)} \lambda_n |e_n,f_n\rangle,
\end{align}
with the Schmidt rank ${r(\Psi)}$, the Schmidt coefficients $\lambda_k>0$ and $\{|e_k\rangle\}_{k=1\dots r}$, $\{|f_k\rangle\}_{k=1\dots r}$ being orthonormal in $\mathcal{H}_1$, $\mathcal{H}_2$, respectively.
Any mixed quantum state $ \rho$ is a convex combination of pure states,
\begin{align}
	 \rho=\sum_k p_k |\psi_k\rangle\langle\psi_k|,
\end{align}
and each vector $|\psi_k\rangle$ of this decomposition has an individual Schmidt rank.
For this distinct decomposition the Schmidt rank of $\rho$ is given by the maximal Schmidt rank of all vectors.
The Schmidt number of the mixed quantum state is given by the minimal Schmidt rank of $\rho$ for all possible decompositions,
\begin{align}
	\nonumber r_{\rm S}( \rho)=\inf\{& r_{\max}: \rho=\sum_k p_k |\psi_k\rangle\langle\psi_k| \\ &\mbox{ and } r_{\max}=\sup_k r(\psi_k) \}.
\end{align}
An overview about the Schmidt number as an entanglement measure can be found in Ref.~\cite{book2}.

For a pure state the Schmidt rank is identical to the Schmidt number, and it counts the minimal number of superpositions of factorizable states needed to generate the state under study.
Therefore, the interpretation in quantum physics is the quantum superposition principle.
In compound Hilbert spaces, this was shown to be the unexpected property in relation to classical physics.
In the following we will throughout use the notion Schmidt number.
The definition of the Schmidt number implies that the suitable LOCC in Definition~\ref{Def:Measure} is the complete set $\mathcal C_{\rm sep}$.
For separable quantum states the Schmidt number is equal to 1.
The simple shift, $r_{\rm S}\to r_{\rm S}-1$, delivers Condition~(i) in the definition for entanglement measures, with $E = r_{\rm S}-1$.

In the following let us consider some basic properties of arbitrary entanglement measures and their connection with the Schmidt number.
It will become clear that the Schmidt coefficients $\lambda_n$ as given in Eq.~(\ref{SchmDec}) play a minor role compared with the Schmidt number for the quantification of entanglement.
The Schmidt number delivers a discrete and monotonous quantification of quantum states with respect to entanglement.

\subsection{Entanglement Invariance under Local Invertibles}

Before quantifying entanglement, it is important to detect the entanglement of a given quantum state.
This can be done by optimized entanglement witnesses~\cite{Witness1,Witness2,SpeVo1}.
These are Hermitian operators $W$ with a non-negative expectation value for separable quantum states.
However, entangles states can have a negative expectation values.

The importance of the above defined LOCC subgroup $\mathcal C_{\rm LI}$ of local invertible operations follows from its influence on the property of entanglement itself.
More precisely, a general LOCC operation can eliminate the entanglement of a quantum state.
However, a general mixed quantum state is entangled, iff it is entangled under local invertibles.

\begin{theorem}\label{Theo:LIent}
	{\bf -- Entanglement under $\mathcal C_{\rm LI}$.}\\
	{\rm (i)} If the operator $W$ is an optimized entanglement witness, then $\left(T_1\otimes T_2\right)^\dagger W \left(T_1\otimes T_2\right)$ is an optimized entanglement witness, with $T_1\otimes T_2$ invertible.\\
	{\rm (ii)} A quantum state 
	\begin{align*}
		\rho'=\frac{\left(T_1\otimes T_2\right)^\dagger \rho \left(T_1\otimes T_2\right)}{{\rm tr}[\left(T_1\otimes T_2\right)^\dagger \rho \left(T_1\otimes T_2\right)]}
	\end{align*}
	is entangled, iff $\rho$ is entangled, with $T_1\otimes T_2$ invertible.
\end{theorem}

The proof is given in~\ref{App:LU}.
These findings of Theorem~\ref{Theo:LIent} are known~\cite{Dehaene,Leinaas}, but in the following we consider these results in the notion of entanglement measures.
This theorem proves that the entanglement of a quantum state is preserved under local operations which are invertible, $\mathcal C_{\rm LI}$.
It delivers also a method to construct an equivalent class of entanglement witnesses detecting the entanglement of arbitrary quantum states.
This can be done in a form that two quantum states, $\rho$ and $\rho'$, share the same kind  of (e.g. bound or free) entanglement, $\rho\cong\rho'$~\cite{Measures}, if they can be transformed into each other by a local invertible transformation,
\begin{align}\label{Eq:SimilarState}
	\rho'=\frac{\Lambda_{\rm LI}(\rho)}{{\rm tr}\,\Lambda_{\rm LI}(\rho)},
\end{align}
with $\Lambda_{\rm LI}\in\mathcal C_{\rm LI}$.
From Theorem~\ref{Theo:LIent} it follows that we should carefully study the quantification of entanglement with respect to $\mathcal C_{\rm LI}$, since this subgroup is of fundamental importance for the property of entanglement itself.

\subsection{Maximally Entangled States and Universal Entanglement Measures}\label{SSec:Proper}
Let $E$ be an arbitrary entanglement measure, and $\rho_{\rm max}$ is maximally entangled,
\begin{align}
	\forall \rho: E(\rho_{\rm max})\geq E(\rho).\label{Eq:maxent}
\end{align}
Now let us define a new entanglement measure,
\begin{align}
	\label{Eq:LocInvEPrime}E'(\rho)\stackrel{\rm def.}{=}E\left(\frac{\Lambda_{\rm LI}(\rho)}{{\rm tr}\,\Lambda_{\rm LI}(\rho)}\right),
\end{align}
with a local invertible operation $\Lambda_{\rm LI}$.
If $E$ is defined by the LOCC $\mathcal C_{\mathcal X}$, then $E'$ is defined by $\mathcal C_{\mathcal X'}$.
An element $\Lambda'$ of $\mathcal C_{\mathcal X'}$ has the form
\begin{align}
	\Lambda'=\Lambda_{\rm LI}^{-1}\circ\Lambda\circ\Lambda_{\rm LI}, \quad \mbox{with} \quad \Lambda\in\mathcal \mathcal C_{\mathcal X}.
\end{align}
Obviously $E'$ is an entanglement measure for $\mathcal C_{\mathcal X'}$, for details see Appendix~\ref{App:ME}.
The state $\rho'_{\rm max}$ satisfies Eq.~(\ref{Eq:maxent}) for the measure $E'$,
\begin{align}
	\rho'_{\rm max}=\frac{\Lambda_{\rm LI}^{-1}(\rho_{\rm max})}{{\rm tr}\,\Lambda_{\rm LI}^{-1}(\rho_{\rm max})}.
\end{align}
However, the initial state $\rho_{\rm max}$ is not necessarily maximally entangled for $E'$.

This new entanglement measure has some surprising properties.
Let us consider the entangled states $\rho=|\phi_r\rangle\langle\phi_r|$, with
\begin{align}
	\label{Eq:Bell}|\phi_r\rangle&=\frac{1}{\sqrt{r}}\sum_{k=1}^r|k,k\rangle,
\end{align}
which are often considered as maximally entangled.
A local invertible transformation is given by $T_1\otimes T_2$,
\begin{align}
	T_1\otimes T_2&=U_1\left(\sum_{k=1}^r \sqrt{r}\lambda_k |k\rangle\langle k|\right)\otimes U_2,\\
	\label{Eq:LocInvLambda}\Lambda_{\rm LI}(\rho)&=(T_1\otimes T_2)\rho(T_1\otimes T_2)^\dagger,
\end{align}
for arbitrary $\lambda_k>0$, and arbitrary unitaries $U_1$ and $U_2$.
Obviously, $T_1\otimes T_2|\phi_r\rangle=\sum_k \lambda_k U_1 |k\rangle\otimes U_2| k\rangle$ is an arbitrary pure entangled state with the Schmidt number $r$ and Schmidt coefficients $\lambda_k$.
A local invertible transformation changes the Schmidt coefficients, whereas the Schmidt number remains unchanged.
From the calculations above, we obtain the following.

\begin{observation}
	For any entanglement measure $E$ with the maximally entangled state $|\psi_{\rm max}\rangle$ exists a measure $E'$, as defined in Eqs.~(\ref{Eq:LocInvEPrime})~and~(\ref{Eq:LocInvLambda}), such that $T_1\otimes T_2|\psi_{\rm max}\rangle$ is maximally entangled for $E'$.
\end{observation}
The state $T_1\otimes T_2|\phi_r\rangle$ is in general not maximally entangled for $E$.
But, this state is maximally entangled with respect to $E'$, and $|\phi_r\rangle$ is not maximally entangled for $E'$.
An exception is an entanglement measure which is invariant under local invertibles.
Thus the question arises which role play the Schmidt coefficients for the quantification?
\begin{definition}\label{Def:uniM}
	An entanglement measure, which is invariant under all local invertibles, elements of $\mathcal C_{\rm LI}$, is called universal entanglement measure.
\end{definition}
Note that such local invertible operations have recently been experimentally demonstrated in the context of noiseless amplification~\cite{Ralph}.
An example for a universal entanglement measure is the Schmidt number $r_{\rm S}$, since $\mathcal C_{\rm sep}$ includes $\mathcal C_{\rm LI}$.
From our observation it follows that the notion maximally entangled is justified only for some special measures $E$, it becomes meaningless in a more general context, here $E'$.


The other way around, any pure state with the same Schmidt number can be considered to be maximally entangled for the universal measure, $E_{\rm uni}$,
\begin{align}
	E_{\rm uni}(\rho)=\sup_{\Lambda_{\rm LI}\in\mathcal C_{\rm LI}} E\left(\frac{\Lambda_{\rm LI}(\rho)}{{\rm tr}\,\Lambda_{\rm LI}(\rho)}\right).
\end{align}
In general, the initial measure $E$ is not a universal entanglement measure.
However, the resulting measure $E_{\rm uni}$ is invariant for pure states with the same Schmidt number but different Schmidt coefficients, the generalization to mixed states is given in Sec.~\ref{SSec:mixed}.
\begin{observation}\label{Obs:Uni}
	If the LOCC, $\mathcal C_{\mathcal X}$, defining the measure $E$ include the set of local invertibles, $\mathcal C_{\rm LI}$, then the entanglement measure is a universal entanglement measure, and $E$ does not depend on the Schmidt coefficients.\hfill$\blacksquare$
\end{observation}
It follows that all similar states $\rho=\rho'$, cf. Eq.~(\ref{Eq:SimilarState}), have the same amount of entanglement.
From the physical point of view it is clear that the number of non-local superpositions (the Schmidt number) is a good characterization of  entanglement.
Since local invertible transformations conserve the Schmidt number, the latter itself may serve as a universal entanglement measure.

In~\cite{Measures} it has been shown for distance-type measures that the idea {\em measure$=$distance$=$amount of entanglement} can be misleading.
The quantification of the amount of entanglement strongly depends on the choice of the distance.
Moreover, the general importance of the local invertible operations becomes clear from Theorem~\ref{Theo:LIent}, proving that the property entanglement of a quantum state is unchanged under local invertibles.


\subsection{Schmidt Rank Monotones}\label{SSec:Proper2}

Let us consider entanglement measures defined by the LOCC $\mathcal C_{\mathcal X}$ which includes local projections.
This is a weak requirement, since many LOCC sub semi-groups fulfill this requirement~\cite{book2}.
The projection $P\otimes \mathbb I_2$,
\begin{align}\label{Eq:Projektor}
	P\otimes \mathbb I_2=\left(\sum_{k=1}^{r-1}|k\rangle\langle k|\right)\otimes \mathbb I_2,
\end{align}
maps the state $|\phi_r\rangle$ to $|\phi_{r-1}\rangle$ (neglecting the normalization).
The LOCC Condition~(ii) in Definition~\ref{Def:Measure} delivers $E(|\phi_r\rangle\langle\phi_r|) \geq E(|\phi_{r-1}\rangle\langle\phi_{r-1}|)$.
In general, a pure state with a given Schmidt number contains less entanglement than a pure state with a higher Schmidt number.
The next section implies the same result for mixed states.
\begin{observation}\label{Obs:Mono}
	An entanglement measure, being defined by LOCC $\mathcal C_{\mathcal X}$ including local projections, is a monotone of the Schmidt number.\hfill$\blacksquare$
\end{observation}
This means that a sequence of quantum states with decreasing number of non-local superpositions, cannot increase its entanglement with respect to $E$.
Thus, the entanglement of Schmidt number states, for a given $r$, delivers upper boundaries for measure with projections and arbitrary quantum states with a Schmidt number less then $r$.

Note the fact that we can also use a deterministic (trace preserving) operation instead of $P\otimes \mathbb I_2$, see Eq.~(\ref{Eq:Projektor}).
Such an operation $\Lambda$ can be given as
\begin{align}
	\Lambda(\rho)=\left(P\otimes \mathbb I_2\right)^\dagger\rho \left(P\otimes \mathbb I_2\right)+\left(|r\rangle\langle r|\otimes \mathbb I_2\right)\rho\left(|r\rangle\langle r|\otimes \mathbb I_2\right).
\end{align}
Obviously follows that $\Lambda(|\phi_r\rangle\langle\phi_r|)$ is a Schmidt number $r-1$ state:
\begin{align}
	\Lambda(|\phi_r\rangle\langle\phi_r|)=\frac{r-1}{r}|\phi_{r-1}\rangle\langle\phi_{r-1}|+\frac{1}{r}|r,r\rangle\langle r,r|.
\end{align}
In the following, we generalize our observations to mixed quantum states.

\subsection{Mixed quantum states, and measuring entanglement measures}\label{SSec:mixed}
So far we have seen properties which are defined by LOCC operations that include local invertible or local projections.
In the following we study entanglement measures including both.
Let us consider an arbitrary mixed quantum state $\sigma_r=\sum_k p_k |\psi_k\rangle\langle\psi_k|$ with a Schmidt number $r_{\rm S}(\sigma_r)=r$.
This state can be generated by an arbitrary pure state $|\tilde\psi_r\rangle$ -- with the Schmidt number $r$ -- and an LOCC operation  $\Lambda\in\mathcal C_{\rm sep}$,
\begin{align}
	\nonumber\sigma_r=&\Lambda(|\tilde\psi_r\rangle\langle\tilde\psi_r|)\\=&\sum_k \left(A_k\otimes B_k\right) |\tilde\psi_r\rangle\langle\tilde\psi_r|\left(A_k\otimes B_k\right)^\dagger\label{Eq:GenLOCC}
\end{align}
with 
\begin{align}
A_k\otimes B_k|\tilde\psi_r\rangle=\sqrt{p_k}|\psi_k\rangle
\label{Eq:mixgen}
\end{align}
as explained above by local invertibles and local projections together with classical mixing.
Let us call the pure state $|\tilde\psi_r\rangle$ the generator of $\sigma_r$.
From the property (ii) of an entanglement measure, we can conclude that for each generator $|\tilde\psi_r\rangle$ of the state $\sigma_r$ holds
\begin{align}
	E(|\tilde\psi_r\rangle\langle\tilde\psi_r|)\geq E(\sigma_r).
\end{align}
Thus we can formulate the following observation.
\begin{observation}\label{Obs:Mixed}
The generator of any mixed quantum state has an equal or a larger amount of entanglement than the generated mixed quantum state.
Under all states $\rho_{\rm max}$ satisfying Eq.~(\ref{Eq:maxent}) must exist a pure state.\hfill$\blacksquare$
\end{observation}
This statement generalizes Observations~\ref{Obs:Uni}~and~\ref{Obs:Mono} to mixed quantum states.

\subsection{Entanglement on average}
Sometimes Condition (ii) in Definition~\ref{Def:Measure} is replaced by a stronger condition
\begin{align}
	\nonumber &{\rm (ii')}\quad \forall \Lambda\in\mathcal C_{\mathcal X}: \rho\stackrel{\Lambda}{\mapsto} \sum_k p_k \rho_k \Rightarrow E(\rho)\geq \sum_k p_k E\left(\rho_k\right).
\end{align}
This condition postulates, that entanglement cannot increase on average~\cite{book2,book3}.
It is straight forward to show that our main Observations~\ref{Obs:Uni}-\ref{Obs:Mixed} remain valid also for the stronger Condition~(ii').

First, let us consider operations of the form $\Lambda(\rho)=[A\otimes B]\rho[A\otimes B]^\dagger$.
Obviously, these operations $\Lambda$ do not create an additional mixture,
\begin{align}
	|\chi\rangle\langle\chi|=[A\otimes B]|\psi\rangle\langle\psi|[A\otimes B]^\dagger,\\
	E(|\psi\rangle\langle\psi|)\geq E\left(\frac{\Lambda(|\psi\rangle\langle\psi|)}{{\rm tr}\,\Lambda(|\psi\rangle\langle\psi|)}\right).
\end{align}
In this case (ii) and (ii') are equivalent.
Thus, the local invertible operation $T$ deliver the same results of ordering of quantum states as concluded in Sec.~\ref{SSec:Proper}. 
The universal measures are independent of the Schmidt coefficients, see Observation~\ref{Obs:Uni} as derived from Definition~\ref{Def:uniM}.
If $\Lambda$ has the form of a local projection $P$, then the monotonic behavior with respect to the Schmidt number in Observation~\ref{Obs:Mono} follows immediately as shown in Section~\ref{SSec:Proper2}.
Thus, the Observations~\ref{Obs:Uni}~and~\ref{Obs:Mono} are also true for (ii').

Starting from the generator $|\tilde\psi_r\rangle$ of the quantum state $\sigma_r$, we obtain an additional mixture, see Eq.~(\ref{Eq:GenLOCC}),
\begin{align}
	\sigma_r=\sum_k p_k |\psi_k\rangle\langle\psi_k|.
\end{align}
It follows from Condition~(ii') that the operation defined in Eq.~(\ref{Eq:GenLOCC}) delivers
\begin{align}
	E(|\tilde\psi_r\rangle\langle\tilde\psi_r|)\geq \sum_k p_k E(|\psi_k\rangle\langle\psi_k|).
\end{align}
Therefore, the results of Observation~\ref{Obs:Mixed} -- the generator $|\tilde\psi_r\rangle$ of the quantum state $\sigma_r$ delivers an upper boundary of the amount of entanglement -- remains valid if we assume (ii') instead of (ii).
In all of our considered cases the condition (ii') delivers the same results as condition (ii).

\subsection{Preliminary results}
So far we have seen that the term maximally entangled state can be used only with respect to a given measure, which led us to the Definition~\ref{Def:uniM} of a universal measure.
We have also shown that entanglement measures have a monotonic behavior with respect to the Schmidt number, see Observation~\ref{Obs:Mono}, and under all maximally entangled states for a given measure must exist a pure state, see Observation~\ref{Obs:Mixed}.
At this point the role of the Schmidt number for arbitrary mixed states as a universal entanglement measure becomes clear.

In general, entanglement measures are defined by a mathematical background.
Not all used LOCC $\mathcal C_{\mathcal X}$ operations can be performed in an experiment.
Thus, the value $E$ is not given by a direct experimental measurement.
The other way around, a general experimental setup cannot use every kind of entanglement, for example bound entangled states for a distillation protocol.
It turns out that the state $|\phi_r\rangle$ -- with equally distributed Schmidt coefficients -- is not the best one to perform quantum computation, see~\cite{GrossEisert}.
Whereas, $|\phi_r\rangle$ is the best state for quantum teleportation of an $r$-qudit.
An entanglement measure optimized for a special experimental setup needs to be found, and we need to find the suitable set of LOCC $\mathcal C_{\mathcal X}$ for this specific experiment.
This measure should quantify the usable amount of entanglement of a quantum state.

Examples of such tasks are the distillation protocols.
In this context, a state that cannot be distilled has the same usable amount of entanglement like a separable quantum state.
This observation leads us to a generalization of the concept of entanglement measures, which will be discussed in the following.

\section{Pseudo-Measures, and Operational Measures}\label{Sec:PM}
The above discussions of entanglement measures in the context of a certain task leads to a generalization of entanglement measures.
Condition (i) in Definition~\ref{Def:Measure} can be relaxed to define pseudo-measures.
\begin{definition}\label{Def:Pseudo}
	The non-negative function $E$ is a pseudo-measure for the LOCC $\mathcal C_{\mathcal X}$, if:
	\begin{align*}
		&{\rm (i')}\quad \sigma \mbox{ separable} \quad \Rightarrow \quad E(\sigma)=0,\\
		&{\rm (ii)}\quad \forall \Lambda\in\mathcal C_{\mathcal X}: E(\rho)\geq E\left(\frac{\Lambda(\rho)}{{\rm tr}\,\Lambda(\rho)}\right).
	\end{align*}
\end{definition}
The new condition (i') implies that even an entangled state can have a vanishing amount of usable entanglement, for a given quantum task to be specified by $\mathcal C_{\mathcal X}$.
Now let us consider the application and usefulness of pseudo-measures.

\subsection{Example: PT entanglement}
For example, let us consider the Peres criterion for the partial transposition (PT)~\cite{Peres}.
A state is entangled, if it does not remain a quantum state under PT, $\rho^{\rm PT}\ngeq 0$.
Entangled states $\rho_{\rm BE}$ with a positive PT are bound entangled states.
These states cannot be used for distillation protocols.
Thus we need to define an entanglement pseudo-measure $E_{\rm PT}$, with
\begin{align}
	\label{Eq:CondPT1}&\rho^{\rm PT}\geq 0 \Leftrightarrow E_{\rm PT}(\rho)=0,\\
	\label{Eq:CondPT2}&\forall \Lambda\in\mathcal C_{\rm distill}: E_{\rm PT}(\rho)\geq E_{\rm PT}\left(\frac{\Lambda(\rho)}{{\rm tr}\,\Lambda(\rho)}\right).
\end{align}
The LOCC $\mathcal C_{\rm distill}$ are the allowed operations for a distillation protocol.
Since all separable quantum states $\sigma$ remain non-negative under PT, it follows $E_{\rm PT}(\sigma)=0$.
However, a state with a non-negative PT is in general not separable, but $E_{\rm PT}(\rho_{\rm BE})=0$.
We conclude that Eqs.~(\ref{Eq:CondPT1})~and~(\ref{Eq:CondPT2}) define the pseudo-measure $E_{\rm PT}$ with respect to Definition~\ref{Def:Pseudo}.
But, it is not an entanglement measure, see Definition~\ref{Def:Measure}.

One possible way for the construction of such a measures is given by measures based on entanglement witnesses, see~\cite{Brandao1,Brandao2},
\begin{align}
	E_{\rm Witness}( \rho)=-\inf_{ W}\{{\rm tr}\,( \rho W)\}=\sup_{W}\{-{\rm tr}\,( \rho W)\},
\end{align}
with entanglement witnesses $W$ of a given form.
A quantum state $\rho$ has a negative PT, if and only if there exists a positive operator, $C=|\psi\rangle\langle\psi|$, with
\begin{align}
	{\rm tr}\,(\rho^{\rm PT}C)={\rm tr}\,(\rho C^{\rm PT})<0.
\end{align}
As we have seen above, any $|\psi\rangle$ can be generated by a $|\phi_r\rangle$ and local invertibles and local projections.
Thus $C^{\rm PT}$ can be generated by a certain $\Lambda\in\mathcal C_{\rm sep}$ and $C^{\rm PT}=\Lambda(V)$, with
\begin{align}
	V = (|\phi_r\rangle\langle\phi_r|)^{\rm PT} = \frac{1}{r}\sum_{k,l=1}^r |k,l\rangle\langle l,k|.
\end{align}
We may define the following entanglement pseudo-measure
\begin{align}
	E_{\rm PT}(\rho)=\sup_{\Lambda\in\mathcal C_{\rm sep}}\left( -\frac{{\rm tr}[\Lambda(\rho)V]}{{\rm tr}\,\Lambda(\rho)}\right).
\end{align}
This pseudo-measure fulfills Eq.~(\ref{Eq:CondPT1}), since a PPT entangled state cannot be distilled.
And, it fulfills Eq.~(\ref{Eq:CondPT2}), since $\mathcal C_{\rm distill}$ is a subset of $\mathcal C_{\rm sep}$.

\subsection{Operational Entanglement Measures}
Let us generalize this situation.
We consider an experimental measurement given by the Hermitian operator $M$.
Now we use the entanglement condition: A quantum state $\rho$ is entangled, iff ${\rm tr}\, \rho M>f_{AB}(M)$~\cite{SpeVo1}, where $f_{AB}(M)$ denotes the maximal expectation value of $M$ for separable states.
We also use the maximal expectation value $f(M)=\sup\{\langle \psi|M|\psi\rangle:\langle \psi|\psi\rangle=1\}$ for all states to define an operational measure $E_M$.
\begin{definition}
	An operational measure $E_M$ is a pseudo measure defined by
	\begin{align}\nonumber
		E_M(\rho)=\sup_{\Lambda\in\mathcal C_{\mathcal X}}\frac{{\rm tr}\,\left(\frac{\Lambda(\rho)}{{\rm tr}\,\Lambda(\rho)} M\right)-f_{12}(M)}{f(M)-f_{12}(M)}.
	\end{align}
\end{definition}
This definition is analogous to the definition of an operational measure for nonclassicality, see~\cite{Gehrke}.

From the experimental point of view we have different devices.
The set $\mathcal X=\{\Lambda_1,\dots,\Lambda_n\}$ denotes the action $\Lambda_k$ of the $k$-th devices onto the quantum state.
Thus, $\mathcal C_{\mathcal X}$ defines arbitrary combinations of the used devices.
The definition of the operational measure $E_{M}$ obviously fulfills the Definition~\ref{Def:Pseudo} for the LOCC given by $\mathcal C_{\mathcal X}$.
For $M=-V$ and $\mathcal C_{\mathcal X}=\mathcal C_{\rm sep}$ we obtain the operational measure for $E_{\rm PT}$,
with $f_{AB}(-V)=0$.

Some measurements do not use any kind of entanglement, e.g. $M=M_A\otimes M_B$.
In this case, $f(M)=f_{AB}(M)$, we define $E_M\equiv 0$.
In general, if the operational entanglement vanishes, $E_{M}(\rho)=0$, then the state is either separable or the setup cannot use the specific kind of entanglement.
The maximally entangled states are states $\rho_{\rm max}$ together with an operation $\Lambda\in\mathcal C_{\mathcal X}$, such that
\begin{align}
	{\rm tr}\,\left(\frac{\Lambda(\rho_{\rm max})}{{\rm tr}\,\Lambda(\rho_{\rm max)}} M\right)=f(M).
\end{align}
The amount of operational entanglement of this state is $E_M (\rho_{\rm max}) =1$.
This is equivalent to the statement, that the state $\rho_{\rm max}$ under the transformation $\Lambda$ is within the range of the maximal eigenvalue of $M$.
Let us summarize these statements in the following proposition.
\begin{proposition} The operational entanglement $E_M$ has the following properties:
	\begin{enumerate}
		\item The operational entanglement is a value between zero and one.
		\item The operational entanglement vanishes, $E_{M}(\rho)=0$, if and only if the state is separable or ($M$,~$\mathcal C_{\mathcal X}$) cannot use the specific kind of entanglement of the state $\rho$.
		\item The operational entanglement is maximal, if and only if the state $\rho_{\rm max}$ under the transformation $\Lambda\in\mathcal C_{\mathcal X}$ is within the range of the maximal eigenvalue of the observable $M$.
	\end{enumerate}
	\hfill$\blacksquare$
\end{proposition}

\section{Summary and Conclusions}\label{Sec:SC}
In conclusion, we have studied the role of local invertible transformations in the context of entanglement.
We have reconsidered the invariance of entanglement under local invertibles, showing that the these transformations are intrinsically related to the property entanglement.
Hence the consideration of local invertibles in the context of entanglement quantification is an important issue.

We have proved that the Schmidt number of a pure state has a larger influence on the amount of entanglement than its Schmidt coefficients.
In particular, the Schmidt number yields a discontinuous entanglement quantification which preserves the requirements of an entanglement measure under the most general class of stochastic separable operations.
To account for this fact, we have defined universal entanglement measures, which are invariant under local invertible transformations and hence independent of the Schmidt coefficients.
Further on, we have shown that a general class of entanglement measures have a monotonic behavior with respect to the Schmidt number.
Since the Schmidt number represents the superposition principle of quantum physics, its general importance for the amount of entanglement has a clear physical background, which is deeply related to the main differences between quantum and classical physics.

For many applications of entangled states it is important to quantify the usable amount of entanglement.
For this purpose, we have considered entanglement pseudo-measures for a certain quantum task, which is specified by the used set of LOCC.
The pseudo-measure is zero, whenever the entanglement of a given quantum state is not useful for the application under consideration.
In terms of pseudo-measures such states are as useful as separable states, even though they may be entangled.

The entanglement pseudo-measures can be further specified as operational measures.
In this case the observables of the experimental setup are included in the definition.
It is important that the observations for entanglement measures are also true for operational measures.
Thus, the Schmidt number gives universal limits for all types of quantum tasks using entanglement as a resource.


\section*{Acknowledgment}
This work was supported by the Deutsche Forschungsgemeinschaft through SFB 652.

\appendix

\section{Local Transformations}\label{App:LU}\label{App:ME}

First of all let us prove Theorem 1 in the presented form.
\paragraph*{Proof of Theorem 1.}
(i) An optimized entanglement witness $W$ is given by: $\langle a,b|W|a,b\rangle\geq0$ and $\langle a_0,b_0|W|a_0,b_0\rangle=0$ for some $|a_0,b_0\rangle=|a_0\rangle\otimes|b_0\rangle$.
Obviously holds
\begin{align*}
	\left(T_1|a\rangle\otimes T_2|b\rangle\right)^\dagger W\left(T_1|a\rangle\otimes T_2|b\rangle\right)&\geq 0,
	\intertext{and}
	\left(T_1|a_0'\rangle\otimes T_2|b_0'\rangle\right)^\dagger W\left(T_1|a_0'\rangle\otimes T_2|b_0'\rangle\right)&= 0,
\end{align*}
with $|a_0'\rangle=T_1^{-1}|a_0\rangle$ and $|b_0'\rangle=T_2^{-1}|b_0\rangle$.
Note that the normalization of the states does not effect relations with zero.\\
(ii) Let $\rho$ be entangled, and $W$ be a optimized witness detecting the entanglement of the state, ${\rm tr}\,\rho W<0$.
It follows for the witness $W'=\left(T_1^{-1}\otimes T_2^{-1}\right) W \left(T_1^{-1}\otimes T_2^{-1}\right)^\dagger$ that
\begin{align*}
	&{\rm tr}\left[W'\frac{\left(T_1\otimes T_2\right)^\dagger \rho \left(T_1\otimes T_2\right)}{{\rm tr}[\left(T_1\otimes T_2\right)^\dagger \rho \left(T_1\otimes T_2\right)]}\right]\\
	=&\frac{{\rm tr}\,\rho W}{{\rm tr}[\left(T_1\otimes T_2\right)^\dagger \rho \left(T_1\otimes T_2\right)]}<0.
\end{align*}
From (i) follows that $W'$ is a optimized witness.
Thus $\rho'$ is entangled.
The opposite direction follows analogously using the inverse local operation.
\hfill$\blacksquare$\\

The set of local unitaries $\mathcal C_{\rm LU}$ is a LOCC operation of any $\mathcal C_{\mathcal X}$.
The condition (ii) delivers for an arbitrary quantum state $\rho$ and an arbitrary local unitary transformation $\rho'=[U_1\otimes U_2]\rho[U_1\otimes U_2]^\dagger$ (together with $\rho=[U_1\otimes U_2]^\dagger\rho'[U_1\otimes U_2]$):
\begin{align}
	E(\rho)\geq E(\rho') \quad \mbox{and} \quad E(\rho')\geq E(\rho).
\end{align}

In Eq.~(\ref{Eq:maxent}) we consider the maximally entangled state $\rho_{\rm max}$, with $E(\rho_{\rm max})={\rm maximal}$.
Now let us prove that the state $\rho_{\rm max}'$ has the same property for the measure $E'$:
\begin{align}
	\nonumber E'(\rho_{\rm max}')=&E\left(\frac{\Lambda_{\rm LI}(\rho_{\rm max}')}{{\rm tr\Lambda_{\rm LI}(\rho_{\rm max}')}}\right)\\
	\nonumber =&E\left(\frac{\Lambda_{\rm LI}\left(\frac{\Lambda_{\rm LI}^{-1}(\rho_{\rm max})}{{\rm tr\Lambda_{\rm LI}^{-1}(\rho_{\rm max})}}\right)}{{\rm tr\Lambda_{\rm LI}\left(\frac{\Lambda_{\rm LI}^{-1}(\rho_{\rm max})}{{\rm tr\Lambda_{\rm LI}^{-1}(\rho_{\rm max})}}\right)}}\right)\\
	\nonumber =&E\left(\frac{\Lambda_{\rm LI}\Lambda_{\rm LI}^{-1}(\rho_{\rm max})}{{\rm tr\Lambda_{\rm LI}\Lambda_{\rm LI}^{-1}(\rho_{\rm max})}}\right)\\
	=&E(\rho_{\rm max})={\rm maximal}.
\end{align}

In addition, we let us now prove that the set $\mathcal C_{\mathcal X'}$ satisfies Condition~(ii) in Definition~\ref{Def:Measure}.
It is obvious that
\begin{align}
	\Lambda_{\rm LI}\circ\Lambda'=\Lambda_{\rm LI}\circ(\Lambda_{\rm LI}^{-1}\circ\Lambda\circ\Lambda_{\rm LI})=\Lambda\circ\Lambda_{\rm LI}.\label{Eq:Trans}
\end{align}
Now we can obtain:
\begin{align}
	\nonumber E'\left(\frac{\Lambda'(\rho)}{{\rm tr}\,\Lambda'(\rho)}\right)=&E\left(\frac{\Lambda_{\rm LI}\circ\Lambda'(\rho)}{{\rm tr}\,\Lambda_{\rm LI}\circ\Lambda'(\rho)}\right)\\
	\nonumber =&E\left(\frac{\Lambda\circ\Lambda_{\rm LI}(\rho)}{{\rm tr}\,\Lambda\circ\Lambda_{\rm LI}(\rho)}\right)\\
	\nonumber =&E\left(\frac{\Lambda\left(\frac{\Lambda_{\rm LI}(\rho)}{{\rm tr\Lambda_{\rm LI}(\rho)}}\right)}{{\rm tr\Lambda\left(\frac{\Lambda_{\rm LI}(\rho)}{{\rm tr\Lambda_{\rm LI}(\rho)}}\right)}}\right)\\
	\leq&E\left(\frac{\Lambda_{\rm LI}(\rho)}{{\rm tr\Lambda_{\rm LI}(\rho)}}\right)=E'(\rho).
\end{align}

\end{document}